%% file: main.tex
\documentclass[runningheads]{llncs}
\usepackage{graphicx}
\usepackage{comment}
\usepackage{xcolor}
\usepackage{hyperref}
\usepackage{svg}
\usepackage{amsmath}
\usepackage{caption}
\usepackage{subcaption}
\usepackage{listings}
\usepackage{color}

\usepackage{soul}
\definecolor{aliceblue}{rgb}{0.94, 0.97, 1.0}
\DeclareRobustCommand{\hllg}[1]{{\sethlcolor{aliceblue}\hl{#1}}}
\PassOptionsToPackage{hyphens}{url}\usepackage{hyperref}
\usepackage{xspace}

\newcommand{\tool}{FMViz\xspace}

\begin{document}
\title{FMViz: Visualizing Tests Generated by AFL at the Byte-level}

%
%
%
%
%

%
\author{
    Aftab Hussain, 
    Mohammad Amin Alipour
}
\institute{
    \email{ahussain27@uh.edu, maalipou@central.uh.edu }\\
    University of Houston, Houston, TX, USA
}

\maketitle              
\input{abstract}

\input{introduction}

\input{related-work}

\input{tool-arch}
\input{demo}

\input{conclusion}

%
%

%
%
%
\bibliographystyle{splncs04}
\bibliography{ref}
\newpage

\appendix

\input{appendix}

\end{document}

%% file: abstract.tex
\begin{abstract}
Software fuzzing is a strong testing technique that has become the de facto approach for automated software testing and software vulnerability detection in the industry.
The random nature of fuzzing makes monitoring and understanding the behavior of fuzzers difficult. 
In this paper, we report the development of Fuzzer Mutation Visualizer (\tool), a tool that focuses on visualizing byte-level mutations in fuzzers. In particular, \tool extends American Fuzzy Lop (AFL) to visualize the generated test inputs and highlight changes between consecutively generated seeds as a fuzzing campaign progresses. 
The overarching goal of our tool is to help developers and students comprehend the inner-workings of the AFL fuzzer better.
In this paper, we present the architecture of \tool, discuss a sample case study of it, and outline the future work.
\tool is open-source and publicly available at \url{https://github.com/AftabHussain/afl-test-viz}.

\keywords{Software Visualization \and Fuzzing \and test input mutation.}
\end{abstract}

%% file: introduction.tex
\section{Introduction}
\label{intro}

Fuzzing has become a widely popular tool for testing programs in the software industry. The overall simplicity of the design of coverage-guided fuzzers (e.g., AFL~\cite{afl}) -- generating test inputs by mutating other test inputs in a pseudo-random fashion while optimizing for code coverage in the test subject, and executing them on test subjects at scale -- has been very effective in finding bugs and vulnerabilities in software systems. Large companies have integrated fuzzers in their testing ecosystem; for instance, Google is continuously running fuzzers on its Chrome browser to find vulnerabilities.

There has been significant research in building efficient fuzzers that can generate interesting test inputs faster in a fuzzing campaign: e.g., grammar-based fuzzing~\cite{gramatron,gram-based-wbf}, data-flow techniques~\cite{tainting,pfuzzer}, stochastic scheduling methods~\cite{mopt}, smarter test selection methods~\cite{seed-select}, etc. Nevertheless, there is one theme that all fuzzers have in common, albeit in varying degrees: randomness, which contributes to the ``black-box" nature of their operation.

This randomness of fuzzers poses a challenge for developers to understand and interpret what operations are being carried out on which test inputs, and reason about the behavior of the fuzzers. While fuzzers, like AFL, do offer high-level statistics on what operations are being performed, the information is shown in a hard-to-follow\footnote{The stats are real-time and constantly-changing.} descriptive manner. They also store data for coverage-increasing test inputs only, and provide no support for understanding how the other tests were generated (which may contain useful information). In addition, the real-time statistics do not portray \textit{which} test inputs are being changed (i.e. mutated). Furthermore, the initial choice of test inputs in a fuzzing campaign can considerably influence its progress~\cite{seed-select}. We thus believe it is necessary to have an approach for understanding how the test inputs are being mutated and address the problem from a visualization angle.  

The importance of software visualization (SV) cannot be over-emphasized. The superiority of visual memory in cognition is discussed in Diehl's seminal work~\cite{diehl-book} where it was mentioned that 75\% of all information from the real world is visually perceived, 13\% through auditory senses, and the rest is perceived through other senses. Despite its value, SV has a huge potential to be realized in software engineering~\cite{Diehl2014PastPA} and, to the best of our knowledge, even more so in fuzzing (we discuss a few existing research we found in Section~\ref{rel-work}). Towards the idea of bringing better visualization in fuzzing, we build a visualization approach for the integral component of almost all state-of-the-art fuzzers: mutation. Our tool, FMViz, helps us see which bytes of a test input undergo mutations during an AFL fuzzing campaign, and thereby makes mutation patterns in fuzzing more perceivable. The tool is light-weight and easy to extend to other fuzzers. We believe this work is a stepping stone in the direction of inspecting the behavior of mutational fuzzers on various test inputs, at a deeper-level. 

\noindent\textbf{Contributions.} The main contributions of this paper are as follows:
 \begin{itemize}
    \item We provide a light-weight approach to visualize fuzzing mutation behavior in AFL by visualizing test inputs generated during fuzzing and highlighting the changes.
    \item We instantiate the approach in \tool by capturing and displaying mutation locations in test inputs that undergo mutation in the fuzzing process -- observing series of FMViz output images help in seeing various mutation patterns that take place during fuzzing.
    \item We present a short demonstration of FMViz with an AFL fuzzing process on \texttt{libxml2}, where we present some mutation patterns captured by the tool.
\end{itemize}

\noindent\textbf{Paper Organization.} In Section~\ref{rel-work}, we present some related literature. In Section~\ref{tool-arch}, we provide details of the architecture of FMViz. In Section~\ref{demo}, we illustrate the results of a demo of FMViz in fuzzing libxml2. We conclude our paper in Section~\ref{concl}, providing future directions.

%% file: related-work.tex
\section{Related Work}
\label{rel-work}

Information visualization has been widely used in different realms of software engineering including bug analysis~\cite{bug-viz}, evolution~\cite{evo-viz}, refactoring~\cite{kw,kcore} -- it makes it easier for developers to understand, analyze, and deploy various software engineering tasks. There are a few works that have adopted visualization techniques in the fuzzing domain. For example, VisFuzz~\cite{Zhou2019VisFuzzUA}, an LLVM plugin that works on top of a modified version of AFL, is an interactive real-time visualization tool that visualizes constraints in the fuzz subject by extracting a call graph and a control flow graph from the subject code. FuzzSplore~\cite{fioraldi2021fuzzsplore} provides statistical visualizations such as a coverage plot, which shows the number of edges that are covered over time by test inputs, and a plot that shows the number of interesting test inputs generated over the campaign. Vainio~\cite{vainio2014use} provides a fuzzing visualization framework that adopts information visualization techniques (e.g. circle packing) to view fuzzing performance data such as CPU and memory usage statistics and power consumption. In~\cite{viz-fuzz-prog}, the coverage of a test subject's call graph is visualized when fuzzed by Hongfuzz~\cite{honggfuzz} and AFL. Unlike our tool, FMViz, none of these works delve into visualizing mutations in fuzzers; FMViz generates visuals of how the mutations are occurring on the test inputs at the byte-level.

%% file: tool-arch.tex
\section{Tool Architecture}
\label{tool-arch}
In this section, we provide an overview of the architecture of \tool.
We also present information on its usage and performance.
 The implementation and documentation of \tool is available in Github\footnote{\url{https://github.com/AftabHussain/afl-test-viz}}.

\begin{figure}
\centering
\begin{minipage}[t]{.6\textwidth}
  \centering
  \includegraphics[scale=0.36]{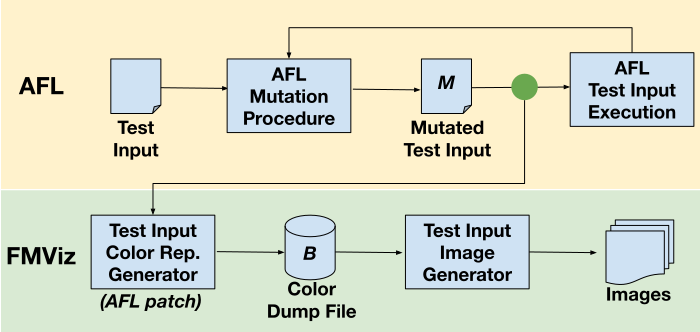}
  \captionof{figure}{Overview of \tool.}
  \label{fig-workflow}
\end{minipage}%
\begin{minipage}[t]{.3\textwidth}
  \centering
  \includegraphics[scale=0.54]{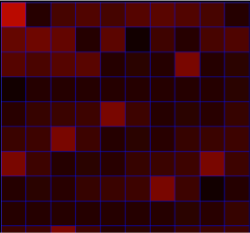}
  \captionof{figure}{Cropped sample image of a test input.}
  \label{fig-sample-op}
\end{minipage}
\end{figure}

Figure~\ref{fig-workflow} depicts the architecture of \tool and its main components, which are: Test Input Color Representation Generator and Test Input Image Generator. 

Figure~\ref{fig-sample-op} shows a sample visualization output of \tool of a test input. 


\subsection{Extracting Color Representations of Test Inputs} 
\tool extends AFL to capture the byte stream representations of new test inputs that are generated as AFL mutates original seeds (Figure~\ref{fig-parts}(a)). 
\tool saves these representations in a single file (a dump file in hexadecimal). Each byte's hex code is chosen to represent a shade of red, depending on the value stored in the byte (we elaborate on the color representation in the following subsection). 
Each line of this file corresponds to the representation of a single test input. 

\subsection{Test Input Image Generator} 
\begin{figure}[htbp]
  \centering
   \includegraphics[scale=0.48]{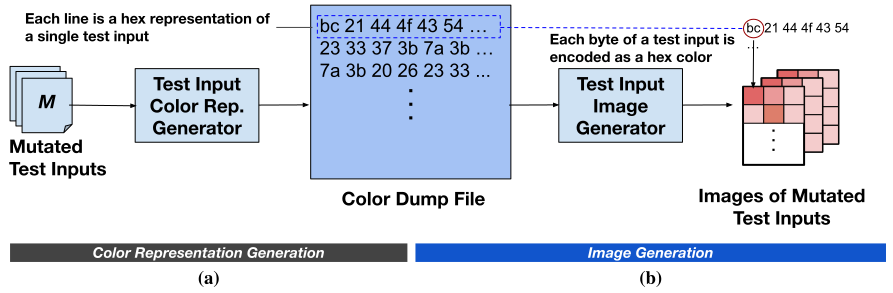}
  \caption{Workflow of the \tool Tool. (a) Test Input Color Representation Generation. (b) Test Input Image Generation}
  \label{fig-parts}
\end{figure}

This piece of our tool (Fig.~\ref{fig-parts}(b)), written in Python, reads the file generated by the color representation generator, line by line, and generates PNG image files (where each line corresponds to a test input as mentioned previously). 
In the image output, each box represents a byte of a test input. 
For obtaining the box colors we use the six-digit hex triplet, a three-byte hexadecimal number, which is typically used for various computing applications, e.g. HTML, SVG, etc., to generate colors. Each of these three bytes show the red, green, and blue components of the color respectively~\cite{hex-trip}. The box color representation for each byte of the test input is evaluated as follows: byte $xy$ is translated to the hex color code $xy0000$, where $xy$ is a hex representation of a test input byte, where $x$ and $y$ each belong to the set of 16 hexadecimal symbols ($0-9$, $A-F$). The box dimensions can be changed to vary the number of test input bytes to display in the image.
The PNG files can be used independently to represent individual test inputs, or can be used to generate a time-lapse video of the evolution of the test inputs using a linear image interpolator~\cite{lin-img-interpolator}, or a screen recorder~\cite{simplescreenrec}.


\subsection{Usage and Performance Considerations}

The present implementation is adapted for an AFL fuzzing campaign with a single test input. Although the overhead of writing to the color dump file is minimal during the fuzzing campaign, since a single file is used, the file can get very large over long fuzzing periods.
We are considering to extend \tool to optimize the storage use through using a more compressed representation of the test inputs.

%% file: demo.tex
\section{Demonstration: Fuzzing \texttt{libxml2}}
\label{demo}

In this section, we present a short demo of \tool. The purpose of this demo is to show how the mutation locations in a test input are visually captured. 

\begin{figure}[htbp]
  \centering
  \includegraphics[scale=0.35]{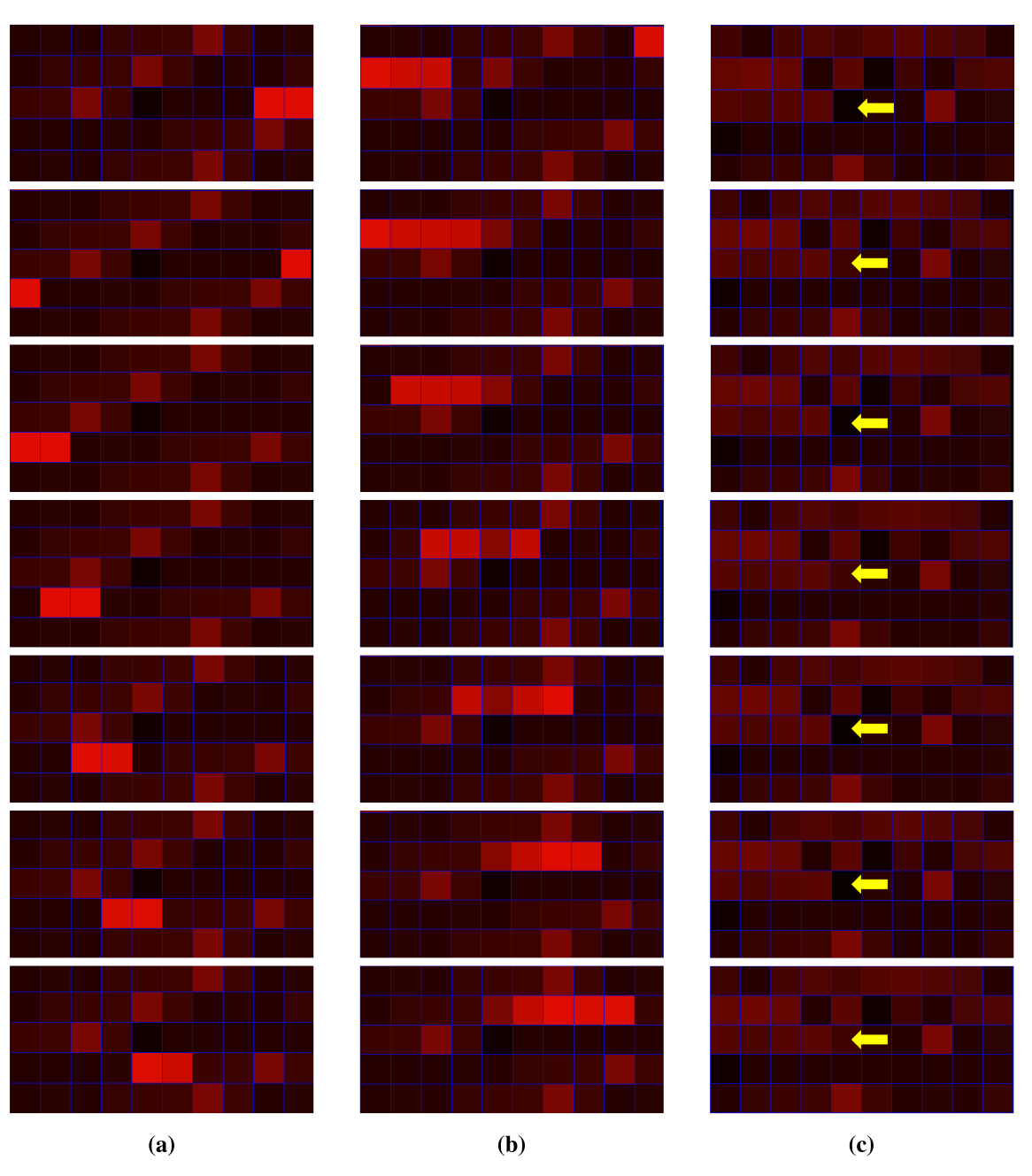}
  \caption{Three observed visual patterns of how an XML test input is mutated during fuzzing \texttt{libxml2} with AFL. (a) 2-byte, shifting mutations (test inputs 7575 to 7581), (b) 4-byte, shifting mutations (test inputs 7851 to 7857), (c) single-byte fixed mutations (test inputs 9358 to 9364) -- yellow arrow added for illustrating changing byte.}
  \label{fig-patterns}
\end{figure}

\subsection{Steps Performed}

We applied \tool's representation generator on top of AFL to fuzz the XML C parser library, \texttt{libxml2}\footnote{We used the version, https://github.com/GNOME/libxml2.git -- commit id. 1fbcf40}\cite{libxml2} for $10$ seconds. This step produced
$9,606$ test inputs and a dump file containing color representations of each of those test inputs. Next, we used \tool's image generator to parse the dump file and generate images for each test. The image generation process took slightly over five minutes for all tests. 
Then viewing a series of consecutive test input color matrix images (in PNG format), using the default system image viewer, revealed patterns (we also produced a time-lapse video from the sequence of images and saved them in a video file, which is available in the repository.). For this experiment, we used a computer system with Intel(R) 1.90GHz Xeon(R) CPU and 64 GB RAM with Ubuntu 18.04.5 LTS.

\subsection{Mutation Patterns Observed}

Figure~\ref{fig-patterns} depicts the mutation patterns that we observed. 
For visualizing each pattern, seven consecutively generated tests are shown. 

\noindent \textit{\textit{2-byte, shifting} mutation pattern} (Figure~\ref{fig-patterns}(a)). Here, in every mutation iteration, the fuzzer mutates a pair of bytes of the test input. This pair-mutation operation progresses by shifting by one byte in the next iteration.

\noindent \textit{\textit{4-byte, shifting} mutation pattern} (Figure~\ref{fig-patterns}(b)). Here, in every mutation iteration, the fuzzer mutates a set of four bytes of the test input. The 4-byte-mutation operation progresses by shifting by one byte in the next iteration.

\noindent \textit{\textit{Single-byte, fixed} mutation pattern} (Figure~\ref{fig-patterns}(c)). Here, in every mutation iteration, the fuzzer mutates the same byte. The changing byte in the figure is shown with the yellow arrow.

%% file: conclusion.tex
\section{Conclusion and Future Work}
\label{concl}

In this work, building on the motivation of software visualization, we presented an easy-to-extend, light-weight visualization tool, \tool, that helps us better perceive the mutation process in the AFL fuzzer. In particular, we visualize bytes of a test input that
undergo mutation during fuzzing. \tool encodes bytes of test inputs as colors and mutations are captured by changes in the colors. 
In the next steps, we plan to augment the visual representation of test inputs by other information such as coverage. We also plan to explore more efficient ways to store representations of test inputs. Furthermore, we plan to evaluate the usefulness of \tool and similar visualization tools in teaching software testing to undergraduate students.

%% file: appendix.tex
\section*{A. Using FMViz}

In this section, we describe the steps to run the first release of FMViz, version 1.0. We use libxml2 as our fuzzing test subject. All commands provided next are for the Linux environment.

\subsection*{Setting up the Environment}

\noindent\textit{1. FMViz Setup}

\vspace{5.5pt}

\noindent In any directory, we clone the FMViz repository as follows\footnote{Currently, the name of the repository has been purposely kept different from FMViz.}:

\vspace{5.5pt}

\noindent\hllg{\texttt{git clone --recursive git@github.com:AftabHussain/afl-test-viz.git}}

\vspace{5.5pt}

\noindent Then we build and install the AFL fuzzer, patched with FMViz's Test Input Color Representation Generator component, by performing the following command:

\vspace{5.5pt}

\noindent\hllg{\texttt{cd afl-test-viz/code/AFL-mut-viz/AFL \&\& make -j32 \&\& make install}}

\vspace{8.5pt}

\noindent\textit{2. \texttt{libxml2} Setup}

\vspace{5.5pt}

\noindent Once we have setup the fuzzer, we build the test subject (\texttt{libxml2}) with AFL's compiler (\texttt{afl-gcc}), which prepares \texttt{libxml2} binaries as fuzzing targets. We first obtain \texttt{libxml2} as follows in a folder outside \texttt{afl-test-viz} directory: 

\vspace{5.5pt}

\noindent\hllg{\texttt{git clone https://github.com/GNOME/libxml2.git \&\& cd libxml2 \&\& git checkout 1fbcf40}}

\vspace{5.5pt}

\noindent Finally, we configure and build \texttt{libxml2} by performing the following command:

\vspace{5.5pt}

\noindent\hllg{\texttt{cd libxml2 \&\& export CC=afl-gcc \&\& ./autogen.sh \&\& make -j32}}

\subsection*{Generating Color Representations of Test Inputs}

We now invoke the first part of FMViz, the augmented AFL fuzzer, which produces color representations (in hex) of test inputs generated while fuzzing the test subject. In this demo, we fuzz the \texttt{libxml2} binary, \texttt{xmllint}. We thus enter the \texttt{libxml2} folder, create an input folder (\texttt{input}), and place in it any XML file as a test input (some sample inputs are available in the FMViz repository):

\vspace{5.5pt}

\noindent\hllg{\texttt{cd libxml2 \&\& mkdir input \&\& cp [path\_to\_xml\_file] input/}}

\vspace{5.5pt}

\noindent Thereafter, we invoke the fuzzer as follows:   

\vspace{5.5pt}

\noindent\hllg{\texttt{export AFL\_SKIP\_CPUFREQ=1 \&\& export LD\_LIBRARY\_PATH=./.libs/ \&\& \\afl-fuzz -i input/ -o output/  -- ./.libs/xmllint -o /dev/null @@}}

\vspace{5.5pt}

\noindent The fuzzing process can be terminated anytime using \texttt{Ctrl+C} -- on termination all results are saved in the output folder, \texttt{output}. Inside this folder, the color dump file \texttt{tests\_generated} contains color representations of all the tests created by the fuzzer. 

\subsection*{Generating Images from Color Representations of Test Inputs}

To generate test input images, we process the color dump file obtained in the previous phase. We place this file along with the Image Generation program (\texttt{viz\_tests.py}) in a separate directory:

\vspace{5.5pt}

\noindent\hllg{\texttt{mkdir process\_color\_rep}}

\vspace{5.5pt}

\noindent\hllg{\texttt{cp libxml2/output/tests\_generated process\_color\_rep/}}

\vspace{5.5pt}

\noindent\hllg{\texttt{cp afl-test-viz/code/viz\_tests.py process\_color\_rep/}}

\vspace{5.5pt}

\noindent Finally we invoke the script:

\vspace{5.5pt}

\noindent\hllg{\texttt{cd process\_color\_rep/ \&\& python viz\_tests.py}}

\vspace{5.5pt}

\noindent The above command generates PNG images for all tests, that are represented in the color dump file, in \texttt{process\_color\_rep} directory:

\vspace{5.5pt}

\noindent\hllg{\texttt{ls | xargs -n 1}}

\vspace{-9.5pt}

\begin{verbatim}
.
.
.
file_000005564.png
file_000005565.png
file_000005566.png
file_000005567.png
file_000005568.png
file_000005569.png
file_000005570.png
file_000005571.png
file_000005572.png
.
.
\end{verbatim}

\begin{figure}[htbp]
  \centering
  \includegraphics[scale=0.36]{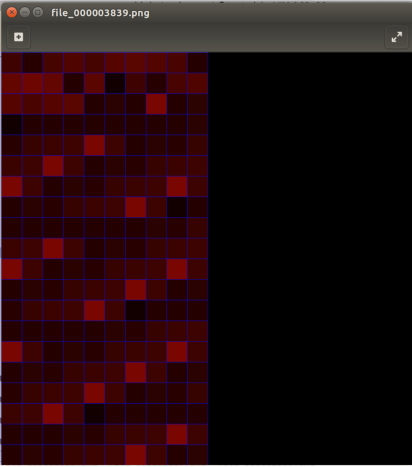}
  \caption{Screenshot of a test input image on Image Viewer}
  \label{fig-imgv-sshot}
\end{figure}

 \noindent A sample screenshot of a test input image, opened with Image Viewer, a default image viewer in Ubuntu, is shown in Figure~\ref{fig-imgv-sshot}. Since the image files for the input tests are named in the order in which they were produced during fuzzing, toggling over consecutive images in the image viewer application shows the trends in mutations. In order to produce a time-lapse video, we use Simple Screen Recorder~\cite{simplescreenrec}, which once installed can be invoked by the command \texttt{simplescreenrecorder} on the terminal. Then by starting recording and toggling over multiple images on Image Viewer by holding the left/right arrow key, we are able to record the mutation transitions that take place.